# Passivity of Metals in the Point Defect Model: Effect of Chloride [Part-III]

Bosco Emmanuel, CSIR-CECRI, Karaikudi-630006, India

**Abstract**

In this paper we study the effect of addition of chloride on the stability of the compact oxide layer pre-existing on a metal surface at a given impressed potential and $pH$ . The variant of the point defect model [PDM] advanced by us recently [1, 2] is used to construct a theory for the chloride-induced build-up of metal vacancies at the metal/film interface and the chloride-induced dissolution of the compact oxide layer. Under the quasi-steady-state approximation the relevant moving boundary value problem is solved exactly and analytical expressions are found for the incubation time, oxide dissolution time, and critical pitting potential, the time evolution of the passive current density and the metal vacancy concentration and the dependence of these quantities on the chloride ion concentration. Some diagnostics are also provided. The replacement of reaction (3) of the original PDM [3] by reaction (3') of the variant [1, 2] is shown to have important consequences for the chloride-ion effect. While the original PDM invoked the Schottky-pair reaction to couple the chloride ion to the cation flux and to the metal vacancy generation by reactions (1) and (4), the present model couples the chloride ion directly to the metal vacancy annihilation by reaction (3') without invoking the Schottky-pair reaction. The anion flux-pinning by chloride is shown to be sufficient to destabilize the oxide layer. An interesting conclusion of the present work is that thicker oxide layers are in general more susceptible to pitting due to the chloride ion and thinner oxide layers destabilize by simple dissolution in the presence of chloride.

## Introduction

In our earlier preprints[1,2] we advanced a variant of the point defect model [1] which rectified a flaw in the original model of Macdonald et al [3] and further elucidated [2] the time dependence of the passive current density and barrier layer thickness during potential switching experiments. In the present paper, which is Part-III of this series, we address the following experiment and work out the corresponding theory based on the corrected version of the point defect model:

- Form the compact oxide layer and a steady passive current density at a given impressed potential V and solution pH in the absence of any aggressive ion like the chloride.
- Add chloride to the electrolyte medium so that the chloride ion concentration switches abruptly to a predetermined value.
- Monitor the transient response of the system in terms of the current density and barrier layer thickness with a view to identify the critical pitting potential and the incubation time

**Theory of Micro-void Generation, Critical Pitting Potential and the Incubation Time**

Following reference [4] we postulate that the following reaction occurs at the film/solution interface and it is at equilibrium with the equilibrium concentration given by equation (2).

$$V_O^{**} + Cl^-(aq) \leftrightarrow Cl_O^* \quad (1)$$

$$C_O(f/s) = \frac{K}{[Cl^-]}\exp\left[\frac{\Delta G_1 - F\phi_{f/s}}{RT}\right] \quad (2)$$

Where $C_O(f/s)$ is the concentration of $V_O^{**}$ at the film/solution interface, $\Delta G_1$ the standard Gibbs free energy of reaction (1), $K$ has the activity corrections, $Cl_o^*$ is a chloride ion occupying an oxygen lattice site and all other symbols have their usual meanings. Unit activity of $Cl_o^*$ is assumed as in [4].

Now the rate of reaction (6) of the point defect model, namely

$$V_O^{**} + H_2O \Rightarrow O_O + 2H^+$$

is pinned at

$$R_6 = k_6.C_O(f/s) \quad (3)$$

with $C_O(f/s)$ given by equation (2). This implies that the flux of $V_O^{**}$ in the barrier layer is pinned by the chloride. In the quasi-steady-state approximation of [2], the rate of reaction (3')

$$m + qV_m + \left(\frac{\chi}{2}\right)O_O \Rightarrow MO_{\chi/2} + \left(\frac{\chi}{2}\right).V_O^{**} + \chi e^{'}$$

will directly be controlled by $R_6$. This flux pinning by chloride has important consequences for micro-void generation and pitting as shown below. Similarly the rates of reactions (1) and (2) will respectively be equal to the rates of reactions (4) and (5) in the quasi-steady-state. Even with the quasi-steady-state approximation, the barrier layer thickness, the metal vacancy concentration and the current density will be seen to be time-dependent.

**Time Evolution of the Metal Vacancy Concentration at the Metal/Film Interface, Incubation Time and Film Dissolution Time**

At the metal/film interface the rate of generation of the metal vacancy $V_m$ by reactions (1) and (2) and its annihilation by reaction (3') are given by

Rate of generation of $V_m$ = $k_2 + k_4$ (4)

Rate of annihilation of $V_m$ = $\frac{q}{(\chi/2)}k_6.C_O(f/s)$ (5)

The net rate of generation of $V_m$ at the metal/film interface is:

$$G_m = (k_2 + k_4) - \frac{q}{(\chi/2)}k_6.C_O(f/s) \quad (6)$$

These metal vacancies diffuse into the metal and we need to solve the diffusion equation with a moving boundary at the metal/film interface. The velocity $v$ of the moving boundary at the metal/film interface at $x = x_B(t)$ is given by

$$v = \frac{dx_B(t)}{dt} = -\Omega.R_3 \quad (7)$$

$$= -\Omega.\frac{k_6.C_O(f/s)}{(\chi/2)} \tag{8}$$

which is a constant for a fixed potential, $[Cl^-]$ and $pH$. The rate of change of the boundary layer thickness $L$ is

$$\frac{dL}{dt} = \Omega\left(\frac{k_6.C_O(f/s)}{(\chi/2)} - R_7\right) \tag{9}$$

$$= R_L \tag{10}$$

$$\therefore \quad L = L_{ss}^0 + R_L.t \tag{11}$$

as $R_L$ is a constant for case A. Note that when $Cl^-$ is added to the system initially at steady state, $R_L$ switches from zero to a negative value. For the quasi-steady-state approximation, the Stephen problem for the diffusion of metal vacancy in the metal can be solved exactly and the solution is

$$C_m(x,t) = C_m(x = x_B(t),t) + A.(1 - \exp\{D_m s^2 t - sx\}) \tag{12}$$

where $s$ is a parameter related to the velocity $v$ of the moving boundary as [5]

$$s = \left(\frac{v}{D_m}\right) \tag{13}$$

and $D_m$ is the diffusion coefficient of the metal vacancy in the metal phase.

$$A = \frac{G_m}{v} \tag{14}$$

The incubation time for pitting may be evaluated by considering the time evolution of the metal vacancy concentration $C_m$ at the moving metal/film interface. Taking the limit $x \to -\propto$ on either side of equation (12), we obtain

$$C_m(x = -\propto, t) = C_m(x = x_B(t),t) + A \tag{15}$$

$$\therefore \quad C_m(x = x_B(t),t) = C_m^{ss} - A \tag{16}$$

where $C_m^{ss}$ is the uniform concentration of the metal vacancy in the metal at the initial steady state. Using equations (6),(8),and (14) in equation (16),

$$C_m(x = x_B(t), t) = C_m^{ss} + \left(\frac{\chi}{2\Omega}\right)\frac{k_2 + k_4}{k_6 C_O (f/s)} - \frac{q}{\Omega} \quad (17)$$

The functional forms for $k_2$ and $k_4$, for case A, are:

$$k_2 = k_2^0.\exp\{a_2 V + c_2 pH\}.\exp\{-b_2 L(t)\} \quad (18)$$

$$k_4 = k_4^0.\exp\{a_4 V + c_4 pH\} \quad (19)$$

Use equations (11), (18) and (19) in (17) to obtain

$$C_m(x = x_B(t), t) = C_m^{ss} - \frac{q}{\Omega} + K_1 + K_2 \exp[-b_2 R_L t] \quad (20)$$

where

$$K_1 = \left(\frac{\chi}{2\Omega}\right)\frac{k_4^0 \exp[a_4 V + c_4 pH]}{k_6 C_O (f/s)} \quad (21)$$

$$K_2 = \left(\frac{\chi}{2\Omega}\right)\frac{k_2^0 \exp[a_2 V + c_2 pH - b_2 L_{ss}^0]}{k_6 C_O (f/s)} \quad (22)$$

As $R_L$ is negative, it is interesting to note that the concentration of the metal vacancy at the metal/film interface is an exponentially increasing function of time with a time constant which may be identified with the incubation time:

$$T_{inc} = \frac{-1}{b_2 R_L} \quad (23)$$

Using equations (9) and (10) and taking the reciprocal of (23)

$$\frac{1}{T_{inc}} = b_2 \Omega R_7 - b_2 \Omega \frac{k_6 \overline{C}_O (f/s)}{(\chi/2)} \frac{1}{[Cl^-]} \quad (24)$$

where

$$\overline{C}_O(f/s) = K\exp\left(\frac{\Delta G_1 - F\phi_{f/s}}{RT}\right) \tag{25}$$

Equation (24) predicts that the reciprocal of the incubation time is linear in the reciprocal of the chloride concentration with a negative slope and positive intercept that depends on the applied potential and $pH$ .

From equation (11) follows a simple equation for the time required for the complete dissolution of the barrier oxide in the presence of chloride:

$$\frac{1}{T_{dis}} = \frac{\Omega R_7}{L_{ss}^0} - \frac{\Omega}{L_{ss}^0}\frac{k_6\overline{C}_O(f/s)}{(\chi/2)}\frac{1}{[Cl^-]} \tag{26}$$

It should be cautioned that this time for complete dissolution of the barrier oxide is not to be confused with the incubation time for pitting which is given by equation (24). Note the functional similarity of equations (24) and (26). The place of $b_2$ in equation (24) is taken by $\frac{1}{L_{ss}^0}$ in equation (26). Clearly $T_{inc}$ will be smaller than $T_{dis}$ if and when $L_{ss}^0$ is larger than $\frac{1}{b_2}$. In the opposite case film dissolution will overtake pitting. As $L_{ss}^0$ increases linearly with the applied potential, pitting will be the failure mode for higher applied potentials and for lower applied potentials chloride-induced film dissolution will be favored. This conclusion is made possible by the fact that $b_2$ is essentially independent of the potential in the point defect model [this independence was assumed in the original PDM and is predicted in the variant of the PDM [1]].

**Critical Pitting Potential**

In the present theoretical framework the critical pitting potential may be identified as the potential above which there is a net metal vacancy generation at the metal/film interface. This condition can be stated as:

$$G_m > 0 \tag{27}$$

Thus the critical pitting potential is the solution of the equation

$$G_m = (k_2 + k_4) - \frac{q}{(\chi/2)} k_6 . C_O(f/s) = 0 \qquad (28)$$

Use the known forms of $k_2$ $k_2$, $k_4$, $k_6$ and $C_O(f/s)$ and solve equation (28) for the critical pitting potential. Note, however, that $k_2$ involves the time-dependent barrier-layer thickness $L(t)$. Just replace $L(t)$ by $L_{ss}^0$ as the system will be critical at a later time $t$ if it is critical initially. Interestingly this statement allows for the possibility that a system which is initially sub-critical may become critical at a later time. Thus a concept of dynamic criticality emerges from the present model. This concept may be visualized by constructing a plot of potential versus time based on equation (28). For each potential, chloride ion concentration and $pH$, there is a time above which criticality sets in. This time is not to be confused with the incubation time or the film-dissolution time but rather marks the initiation of the process which terminates at the incubation time or the film-dissolution time. For oxides that are anionic conductors equation (28) may be further simplified as

$$k_2 - \frac{q}{(\chi/2)} k_6 . C_O(f/s) = 0 \qquad (29)$$

After inserting the relevant functional forms and simplifying, there results

$$V_{PIT} = \bar{V}_{PIT} - \frac{2.303 \log[Cl^-]}{\left(a_2 - a_6 + \frac{\alpha F}{RT}\right)} \qquad (30)$$

where

$$\bar{V}_{PIT} = \frac{2.303 \log\left[\frac{qKk_6^0}{(\chi/2)k_2^0}\right] + b_2 L_{ss}^0 + \frac{\Delta G_1}{RT} - \frac{F\phi_{f/s}^0}{RT} + (c_6 - c_2 - \frac{\beta F}{RT}) pH}{\left(a_2 - a_6 + \frac{\alpha f}{RT}\right)} \qquad (31)$$

Note we have set $L(t) = L_{ss}^0$.

**Passive Current Density Evolution on the Addition of Chloride**

The general expression for the passive current density is given by

$$i = F.\{\delta(k_2 + k_4) + 2.J_O + (\delta - \chi).R_7\} \tag{32}$$

After inserting the relevant functional forms we obtain

$$i = F.\{\delta.\bar{k}_2 \exp\{-b_2 L_{ss}^0\}\exp\{-b_2 R_L t\} + \delta.\bar{k}_4 + 2\frac{\bar{J}_O}{[Cl^-]} + (\delta - \chi).\bar{k}_7 C_R^n\} \tag{33}$$

where the potential and $pH$ dependent quantities are:

$$\bar{k}_2 = k_2^0 \exp\{a_2 V + c_2 pH\} \tag{34}$$

$$\bar{k}_4 = k_4^0 \exp\{a_4 V + c_4 pH\} \tag{35}$$

$$\bar{k}_7 = k_7^0 \exp\{a_7 V + c_7 pH\} \tag{36}$$

$$\bar{J}_O = k_6^0 K \exp\left[\frac{(a_6 - \alpha F)V + (c_6 - \beta F)pH + \Delta G_1 - F\phi_{f/s}^0}{RT}\right] \tag{37}$$

As $R_L$ is negative the current density will rise with time exponentially on the addition of chloride to the medium. This rise will continue till the time of complete dissolution of the oxide film or pitting whichever is earlier. In addition equation (33) is not expected to hold very near to zero time as the quasi-steady-state will take some time to establish.

**Conclusions**

In [4] Macdonald et al resorted to the Schottky-pair creation mechanism in order to relate the action of chloride at the film/solution interface on the metal vacancy generation/annihilation at the metal/film interface. This was necessitated by the fact that the reaction (3) in the original point defect model missed out the metal vacancy annihilation. However there is no real need to invoke Schottky-pair creation mechanism as reaction (3’) of the corrected PDM naturally couples the action of chloride at the film/solution interface and the metal vacancy annihilation at the metal/film interface.

The defect reactions considered in the present paper are:

Metal | Barrier Layer | Porous Layer OR Solution

(1) $m + V_M^{\chi'} \Rightarrow M_M + V_m + \chi.e'$

(2) $m + V_{M_i} \Rightarrow M_i^{\chi+} + V_m + \chi.e'$

(3') $m + qV_m + \left(\frac{\chi}{2}\right) O_O \Rightarrow MO_{\chi/2} + \left(\frac{\chi}{2}\right) .V_O^{**} + \chi e'$

(4) $M_M \Rightarrow M^{\delta+}(aq) + V_M^{\chi'} + (\delta - \chi).e'$

(5) $M_i^{\chi+} \Rightarrow M^{\delta+}(aq) + V_{M_i} + (\delta - \chi).e'$

(6) $V_O^{**} + H_2O \Rightarrow O_O + 2H^+$

(7) $MO_{\chi/2} + \chi H^+ \Rightarrow M^{\delta+}(aq) + \left(\frac{\chi}{2}\right) H_2O + (\delta - \chi)e'$